\documentclass[referee]{raa}
\usepackage{graphicx,times}
\usepackage{natbib}
\usepackage{amssymb,amsmath}
\bibpunct{(}{)}{;}{a}{}{,}

\usepackage[a4paper=true,dvipdfm=true,pagebackref=true,unicode]{hyperref}
\hypersetup{pdftitle = The title of my PDF, pdfauthor = My name, pdfsubject= The subject, pdfkeywords = keyword1 keyword2 keyword3}
\hypersetup{colorlinks = true, linkcolor = black, anchorcolor = black, citecolor = black, filecolor = black, pagecolor = black, urlcolor = black}

\begin{document}

   \title{Investigation of the possible source for solar energetic particle event of 2017 September 10}

 \volnopage{ {\bf 20XX} Vol.\ {\bf X} No. {\bf XX}, 000--000}
   \setcounter{page}{1}

   \author{Ming-Xian Zhao\inst{1}, Gui-Ming Le\inst{1,2}, Yu-Tian Chi\inst{3}
   }

   \institute{Key Laboratory of Space Weather,
              National Center for Space Weather,
              China Meteorological Administration,
              Beijing, 100081, China; ({\it Legm@cma.gov.cn})\\
        \and
             School of Computer Science, Anhui University of Technology,
             Maanshan 243032, China\\
	    \and
             School of Earth and Space Sciences,
             The University of Science and Technology of China,
             Hefei, Anhui, 230026, China\\
\vs \no
   {\small Received 20** June **; accepted 20** July **}
}

\abstract{According to the solar protons' data observed by Geostationary Operational Environmental Satellites (GOES) and neutron monitors on the ground of the Earth, and near-relativistic electrons data measured by the ACE spacecraft, the onset times of protons with different energies and near-relativistic electrons have been estimated and compared with the time of solar soft and hard X-ray and radio burst data. The results show that first arriving relativistic and non-relativistic protons and electrons may be accelerated by the concurrent flare. The results also suggest that release times of protons with different energies may be different, and the protons with lower energy may release earlier than those with higher energy. Some protons accelerated by concurrent flares may be further accelerated by CME-driven shock.
}

   \authorrunning{M.-X. Zhao, G.-M. Le \& Y.-T. Chi}            %author_head in even pages
   \titlerunning{Solar energetic particle event of 2017 Sep 10}  % title_head in odd pages
   \maketitle

%________________________________________________ sections below
%
\section{Introduction}           %% first-level sections will be auto-capitalized
\label{sect:intro}

A large gradual SEP event is often accompanied with both a gradual flare and coronal mass ejection (CME). Whether gradual flare contributes to the production of protons in a large gradual SEP event is still a open question. The results of some papers suggested that relativistic solar protons (RSPs) may be accelerated by concurrent flares (e.g., Aurass et al. 2006; Bazilevskaya 2009; Grechnev et al. 2008, 2015; Klein et al. 2014; Kouloumvakos et al. 2015; Le et al. 2006, 2013, 2014; Li et al. 2007a, 2007b, 2009; Masson et al. 2009; Miroshnichenko et al. 2005a; Perez-Peraza et al. 2009; Simnett 2006). A simple and effective method is to calculate the release time of RSPs and compares it with the time of concurrent flare and metric Type II radio burst, and then the solar origin of the first arriving particles can be judged. The release times for many ground level enhancement (GLE) events inferred from velocity dispersion analysis (VDA) seemly supported that RSPs may be accelerated by CME-driven shock (Reames 2009a, 2009b). The path length for protons propagating from the Sun to the Earth inferred by VDA is usually much longer than the nominal Park spiral line. To be noticed that the release time for some GLE event obtained by different researchers may be different. For example, the solar origin of the RSPs in the SEP event of 2003 October 28 obtained by Miroshnichenko et al. (2005a) is different from the one obtained by Reames (2009a). The possible sources for  non-relativistic protons have also been investigated, the results showed that E$>$30 MeV protons may be mainly accelerated by the concurrent flares for SEP events with source location in the longitudinal area ranged from W40 to W70 (Le et al. 2017a, 2017b). When a large gradual SEP happened, near relativistic electrons and even relativistic electrons are often accompanied by long lasting and intense type III burst. The long lasting and very intense type III burst is termed as type III-l burst(Cane et al. 2002). The origins of near relativistic electrons and relativistic electrons have been extensively investigated and different researches have different points of view on the origin of the near-relativistic electrons. (e.g., Cane et al. 2002, 2006, 2010; Cane 2003; Kahler 2007; Cliver et al. 2009 and reference therein). The conclusion that near relativistic electrons or higher energy electrons are shock acceleration are based on the assumption that the path length traveled by near relativistic electrons from the Sun to the Earth is 1.2 AU.

An X8.2 flare of 2017 September 10 was accompanied by very fast CME. Protons escaped from the Sun with energies from keV to GeV and near relativistic electrons were observed. According to the RSPs data observed by neutron monitors (NMs) and GOES spacecraft, and the near-relativistic electrons observed by the ACE spacecraft, as well as solar soft and hard X-ray and radio burst data, solar release times (SRTs) of protons and electrons with different energies that occurred on 2017 September 10 will be estimated and then compared with the flare time and the metric type II radio burst onset time to speculate the possible solar source for protons with different energies. This is the motivation of this paper. Data observation is presented in Section \ref{sect:Obs}. Data analysis is presented in Section \ref{sect:data}. Discussion and Summary is presented in Section \ref{sect:discussion}.

\section{Observation data}
\label{sect:Obs}

\begin{figure}
   \centering
   \includegraphics[width=10.0cm, angle=0]{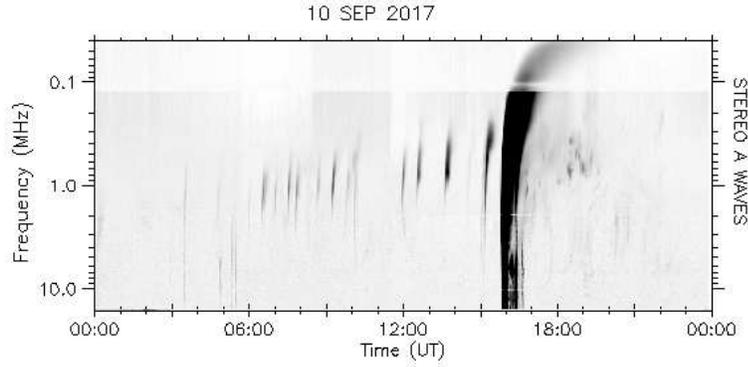}
   \caption{Type III radio burst associated with the large SEP event of 2017 September 10.}
   \label{fig:typeIII}
\end{figure}

\begin{figure}
   \centering
  \includegraphics[width=10.0cm, angle=0]{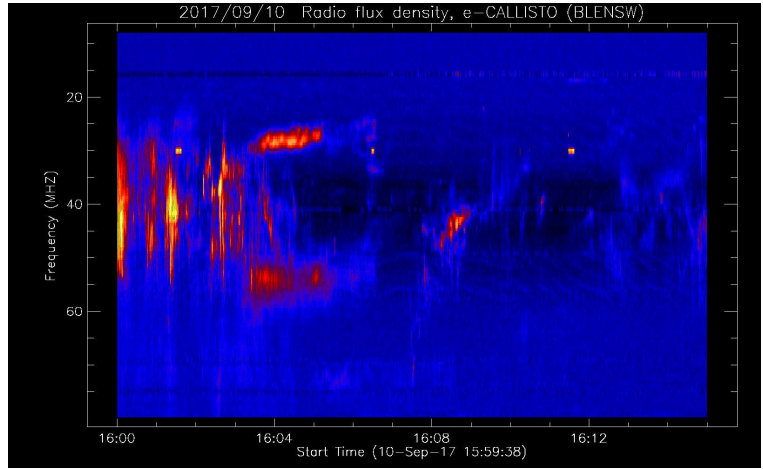}
   \caption{Metric Type II radio burst associated with SEP event of 2017 September 10.}
   \label{fig:typeII}
\end{figure}

Active region (AR) 12673 located at S08W88 produced a X8.2 flare. This flare started at 15:35 UT, peaked at 16:06 UT and ended at 16:31 UT, 10 September 2017. The onset time of metric Type III observed by Stereo A is 15:48 UT shown in Figure \ref{fig:typeIII}. Hard x-ray (HXR) with energy 100-300 keV peaked at 15:56 UT, while HXR with energy 300-1000 keV peaked at 15:59:30 UT, 10 September 2017. The flare was accompanied by a very fast CME. The CME entered LASCO C2 and C3 view field at 16:00 UT and 16:06 UT, 10 September 2017 respectively. The onset time of Type II radio burst generated by the shock driven by the fast CME is slightly later than 16:03 UT, 10 September 2017, which is obtained from e-Callisto (http://e-callisto.org/) shown in Figure \ref{fig:typeII}.

\begin{figure}
   \centering
  \includegraphics[width=10.0cm, angle=90]{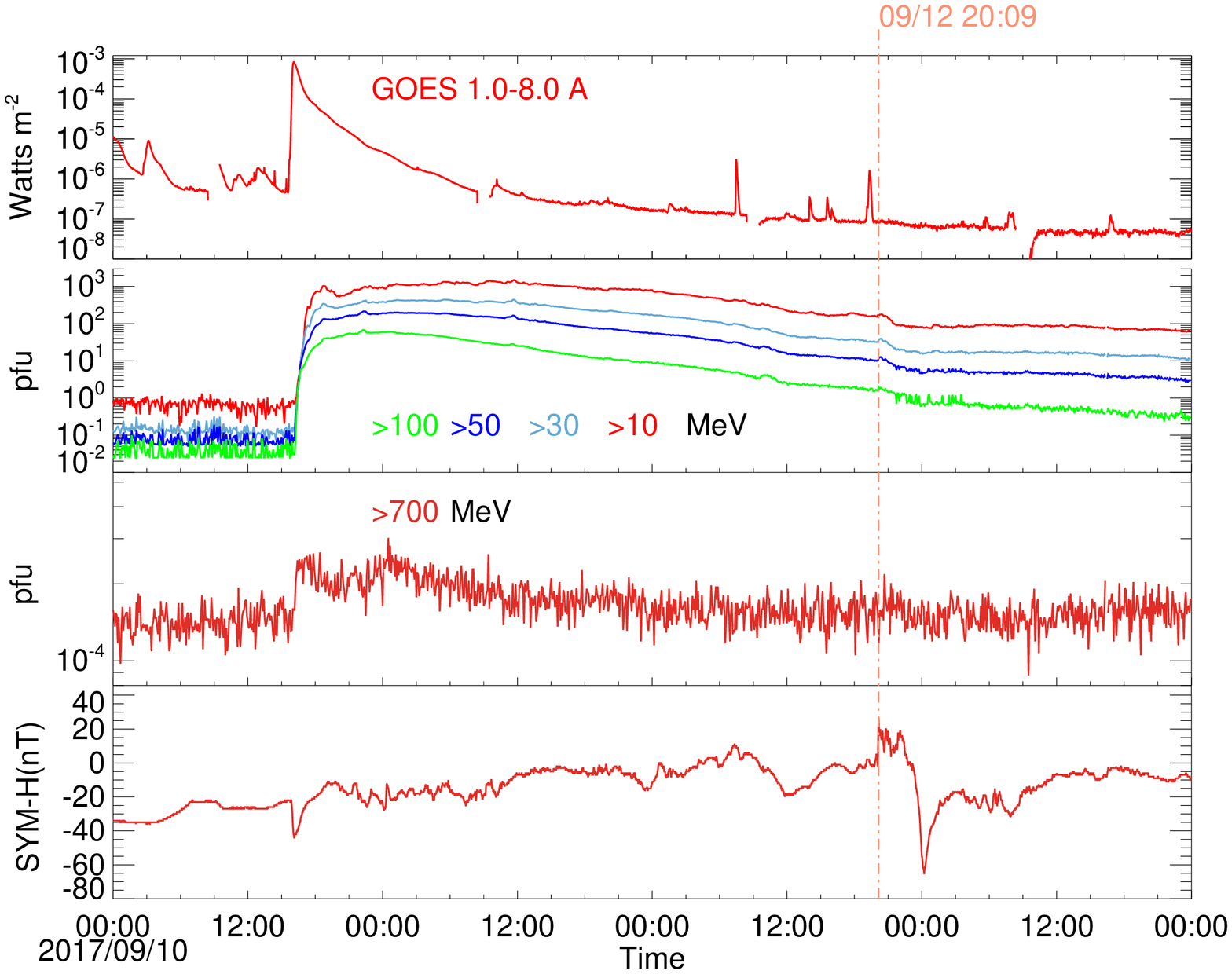}
   \caption{The SXR emission and the fluxes of protons with different energies varied with time on 2017 September 10. From the top to bottom, it indicates the flux of SXR in 1-8~\AA,the flux of protons with different energies ($E>10$ MeV, $30$ MeV, $50$ MeV and $100$ MeV protons), the flux of $E>700$ MeV protons, and the 1-minute time resolution SYM-H index.}
   \label{fig:SXR}
\end{figure}

The flux of protons with different energies observed by GOES increase very quickly after X8.2 flare shown in Figure \ref{fig:SXR}.

\begin{figure}
   \centering
  \includegraphics[width=10.0cm, angle=90]{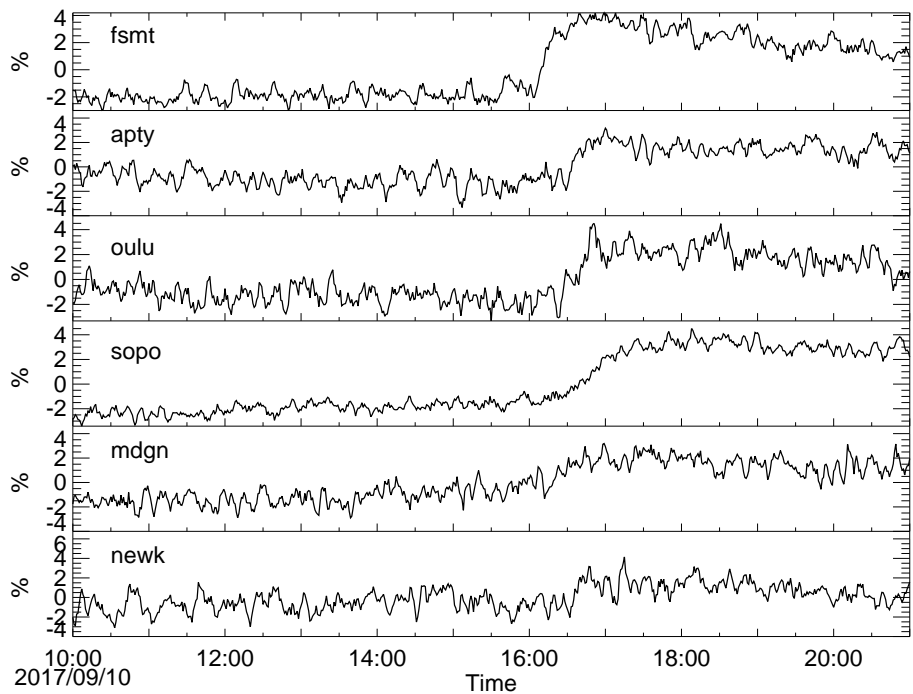}
   \caption{The GLE event observed by several NMs on 2017 September 10.}
   \label{fig:GLE}
\end{figure}

A standard deviation, $\delta$, calculated by the formula below, will be used to estimate the onset time of first arriving particles:
\begin{equation}\label{eq:01}
\delta = \sqrt{\sum_{i=1}^{N}\left(x_i-\overline{x}\right)^2/N}
\end{equation}
where $x_i$ and $\overline{x}$ are the real and averaged value during the quiet period. $N$ is the number of measurements (experimental records) during quiet period. The estimation of the onset time for first arriving particles is based on the intensity of particle exceeds $\overline{x}+2\delta$(e.g., Mewaldt et al. 2003; Tylka et al. 2003; Miroshnichenko et al. 2005b). Using this criterion, we can obtain the onset times of first arriving particles from the data observed by GOES or by other instruments.
\begin{equation}\label{eq:02}
t_{s}=t_{o}-L/v
\end{equation}
where $t_{s}$ is the release time on the Sun, while $t_{o}$ is the onset time of particles observed in situ, such as observed by GOES or by the NMs on the ground.

The onset time estimated for $E>700$ MeV protons is 16:15 UT$\pm$5 min, while the onset time estimated for $E>10$ MeV protons is 16:25 UT$\pm$5 min. However, the onset times estimated for $E>30$ MeV protons, $E>50$ MeV protons and $E>100$ MeV protons are 16:20 UT$\pm$5 min.

The GLE event caused by the interaction between RSPs and atmosphere were observed by several
neutron monitors (NMs) shown in Figure \ref{fig:GLE}. According to formula (\ref{eq:01}), the onset time for each NM shown in Figure \ref{fig:GLE} has been estimated and listed in Table \ref{tab:SEPtime}. In fact, all NMs at Earth surface with geomagnetic cutoff rigidities below 1 GV have a same cutoff (about 1 GV), which is determined by the atmosphere absorption only(e.g., Miroshnichenko 2001; Biber 2002). The different arriving times for RSPs suggest that there was a very strong anisotropy in RSPs at the early phase of the SEP event. The onset time registered by FMST NM with cutoff rigidity 1.0 GV is earlier than that registered by MDGN NM with cutoff rigidity 2.09 GV, suggesting that RSPs with higher energy may release later than RSPs with lower energy. We can also see from Figure \ref{fig:GLE} that the enhancement of cosmic rays observed by Newk, which has a cutoff rigidity 2.4 GV, is not obvious, suggesting that few protons have energy higher than than 1.64 GeV.

\section{Data analysis}
\label{sect:data}
\subsection{Estimation of the path length traveled by SEPs}
\label{sect:sep}
\begin{figure}
   \centering
  \includegraphics[width=10cm, angle=90]{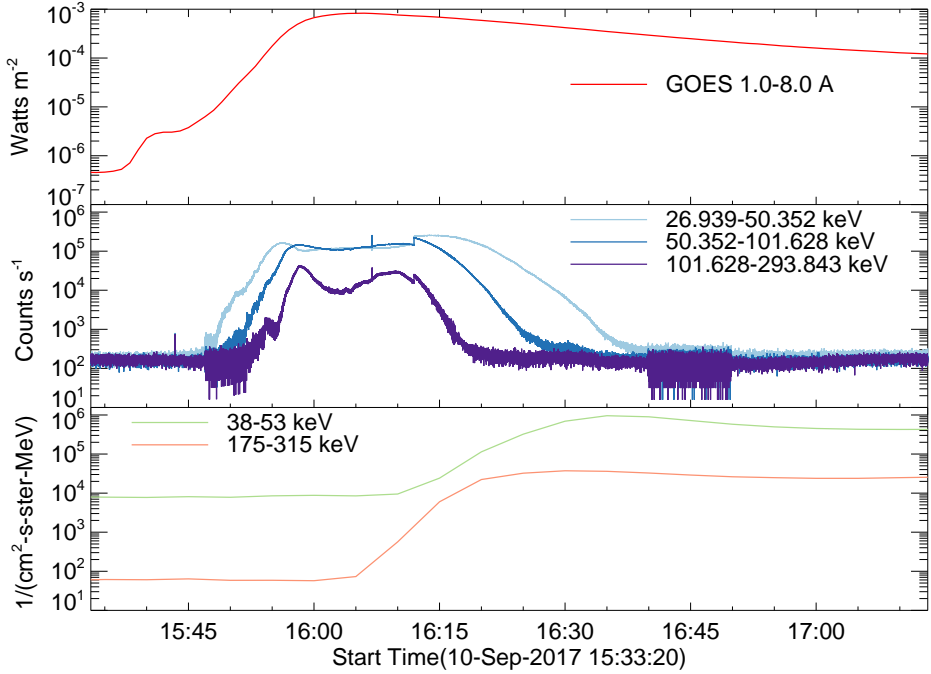}
   \caption{The SXR emission and the fluxes of electrons with two energy channels. From the top to bottom panel, it indicates the flux in 1-8\AA, hard X-ray flux in three channels and the flux of near relativistic electrons in two channels respectively.}
   \label{fig:HXR}
\end{figure}

Type III burst are caused by streams of electrons with energy lower or higher than 25 keV. The presence of metric type and lower frequency type III burst indicates that the field lines over the associated active region have been opened and the particles escape into the interplanetary space. The onset time of type III radio burst shown in Figure \ref{fig:typeIII} is 10 minutes later than the start time of X8.2 flare. The type III radio burst shown in Figure \ref{fig:typeIII}, which lasted tens of minutes, is a typical type III-l burst. Different people have different points of view on the origin of type III-l burst and their relationship with large gradual SEP events (e.g., Cane et al. 2002, 2006, 2010; Cane 2003; Cliver et al. 2009; Kahler 2007 and reference therein). The start time of type III-l burst shown in Figure \ref{fig:typeIII} is 15:48 UT, 15 minutes earlier than the start time of type II burst shown in Figure \ref{fig:typeII}, suggesting that it is impossible that the electrons responsible for the type III-l burst were accelerated by CME-driven shock. The release times of near relativistic electrons inferred from VDA are almost always delayed relative to the start times of associated type III-l radio burst and even in the small electron events in which electrons were believed to be accelerated by the concurrent flares (Cane and Lario 2006). It has been known that electrons can be efficiently accelerated in solar flares (Lin et al. 2003). The start time of type III-l burst can be treated as the solar release time of electrons. Based on the time difference between the start time of type III-l burst and the onset time of 38-53 keV electrons observed by the spacecraft ACE, the path length traveled by electrons with energy 38-53 keV electrons is estimated as $\sim1.7$$\pm$0.22 AU. According to the path length $\sim1.7$$\pm$0.22 AU, the solar release time of near-relativistic electrons with energies 173-315 keV was about 15:48:42 ST$\pm$5 minutes, which is consistent with the first peak time of 100-300 keV HXR shown in Figure \ref{fig:HXR}. The fluxes of HXR in three channels shown in Figure \ref{fig:HXR} are obtained from the FERMI Gamma Ray Burst Monitor(Meegan et al. 2009).

\subsection{Estimated CME and CME-driven shock speed}
\label{sect:cme}

By using coronagraph images taken by SECCHI/COR2 instruments on board STEREO-A and STEREO-B spacecraft and Large Angle and Spectrometric Coronagraph (LASCO) suite (Brueckner et al. 1995) aboard the Solar and Heliospheric Observatory(SOHO) , a three dimensional CMEs are allowed to be modeled with the Graduated Cylindrical Shell (GCS) model  (Thernisien 2006,2009) from low coronal to 1 AU. As the communications to STEREO- B were interrupted since Oct 1, 2014, so we only use coronagraph images from SOHO/LASCO and STEREO-A to reconstruct this CME. The shape of GCS flux rope model is reminiscent of a hollow croissant and can be parameterized with six free parameters. These six parameters are carrington longitude $\Phi$, heliospheric latitude $\theta$, tilt angle $\gamma$, the leading front edge distance of the tracked structure from the Sun $h_{front}$, the half angle of the shell $\alpha$, and the aspect ratio $\kappa$. There is also a spheroid model to reconstruct the shock with six free parameters, longitude, latitude, tilt angel, height, and the major and minor axis of the spheroid($\varepsilon$ and $\kappa$).

\begin{figure}[ht]
  \begin{minipage}[t]{0.4\linewidth}
  \centering
   \includegraphics[width=7.5cm]{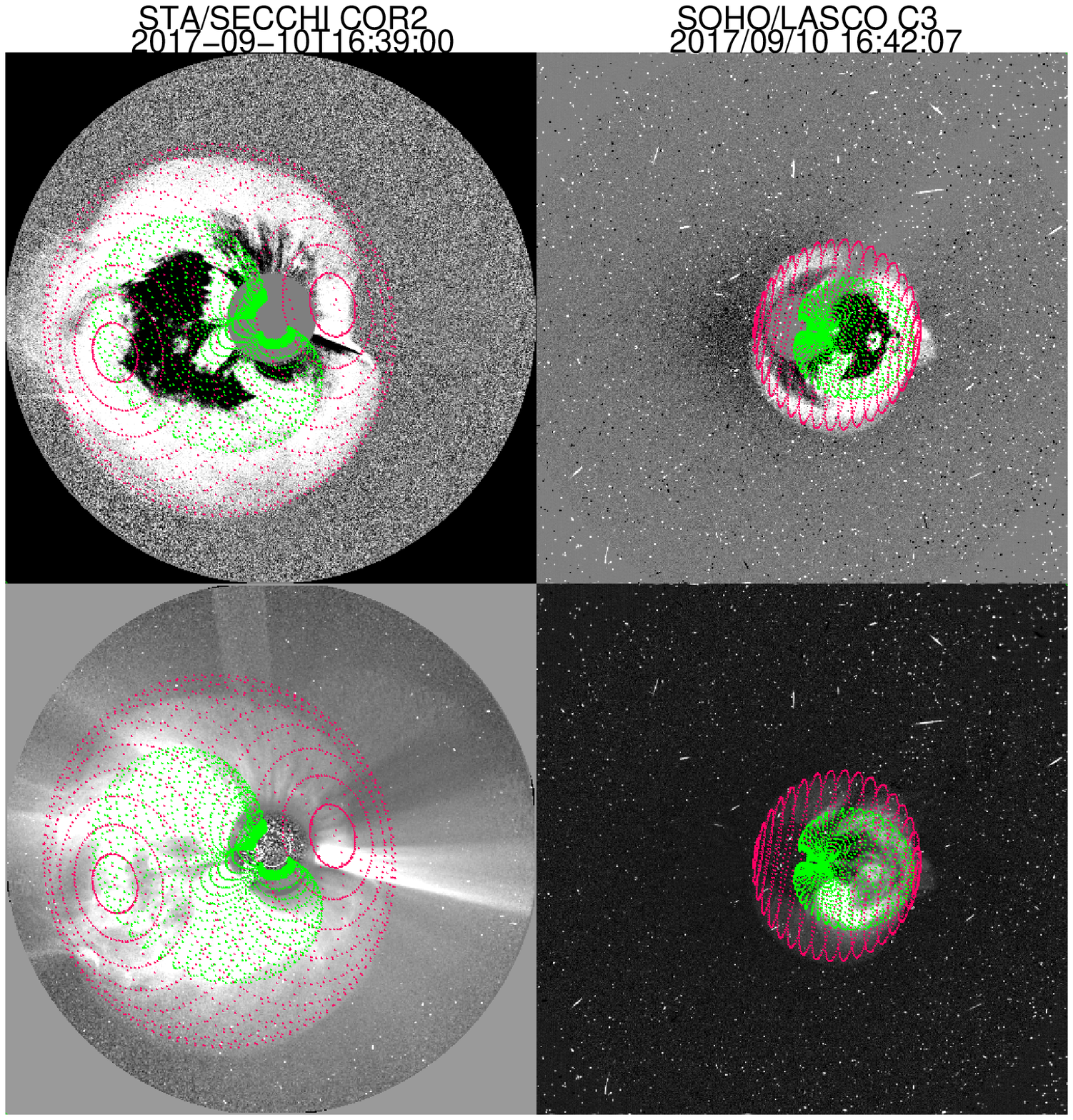}
  \end{minipage}
  \begin{minipage}[t]{0.62\textwidth}
  \centering
   \includegraphics[width=8.0cm]{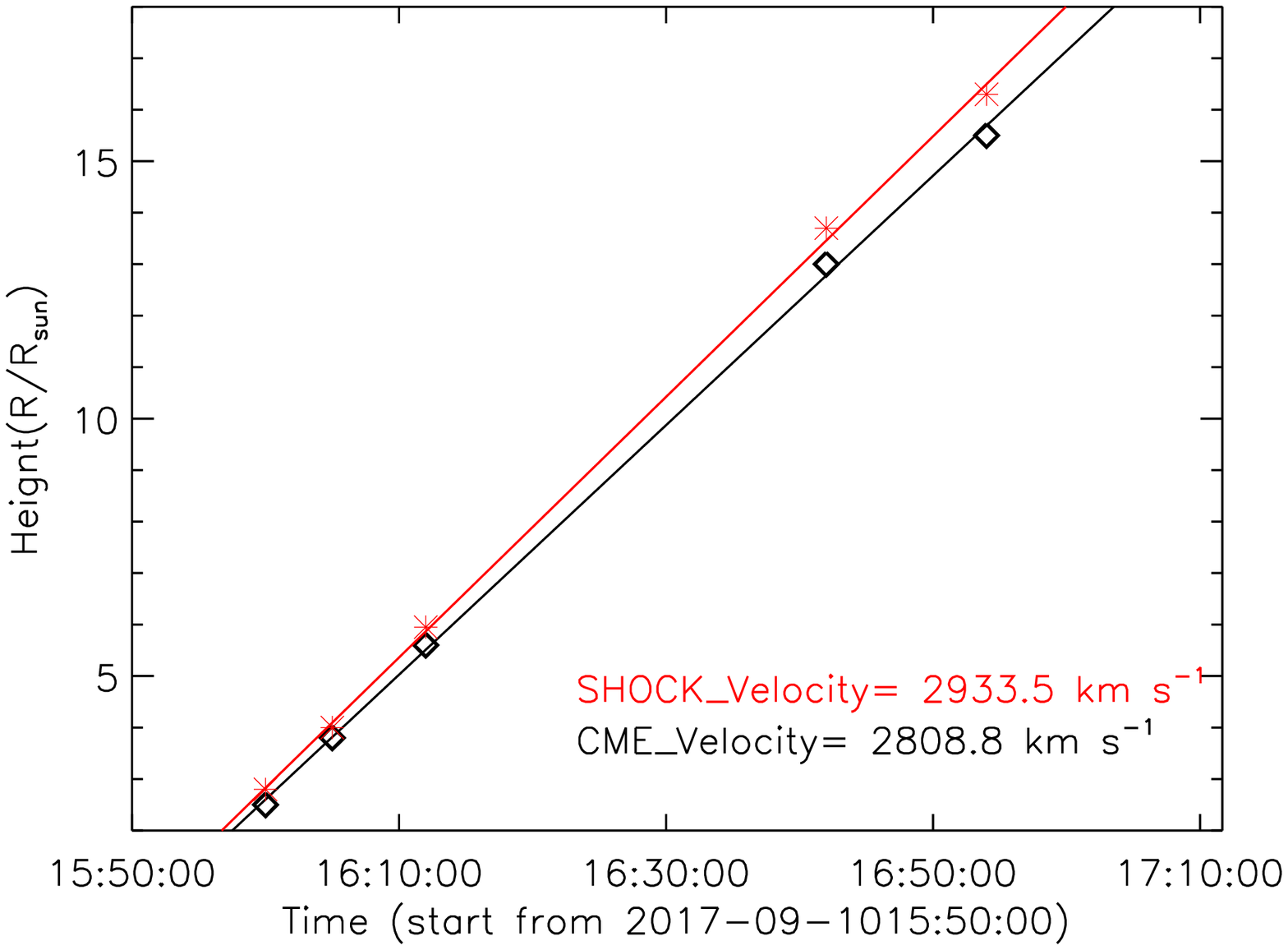}
  \end{minipage}
  \caption{Model fitting of CME front and shock front. The green mesh shows the GCS fitting to the CME, and the red mesh shows the spheroid fitting to the shock front. The linear fit of CME front height(black) and shock front height(red) to time is shown in the right image.}
  \label{fig:CME}
\end{figure}

In this article, we focus on the velocity of the CME and shock, so the height of flux rope  front and the height of shock front are mainly concerned. The CME flux rope structure can be seen more clearly in direct images (Figure \ref{fig:CME} bottom panels), while the shape of shock are more clearly in running difference images (Figure \ref{fig:CME} top panels). As shown in Figure \ref{fig:CME} left four panels,  the red mesh (spheroid model) overlaid shows the CME shock front and the green mesh (GCS model) shows the CME flux rope front.
After success fitting of flux rope model and spheroid shock model for a series of COR2 and C3 images at different times, a set of height-time data will be obtained. All fitting parameters of the GCS model are  in Table \ref{tab:GCSfit} and the spheroid model parameters of shock are in Table \ref{tab:Shockfit}. The velocity of CME flux rope front and shock front can be derived after the linear fitting of height and time. The height-time fitting results are shown in Figure \ref{fig:CME}. The fitted CME speed is about 2808.8 km/s. while CME-driven shock speed is around 2933.5 km/s.

\begin{table}
\bc
\begin{minipage}[]{110mm}
\caption[]{GCS model parameters of CME at different time\label{tab:GCSfit}}\end{minipage}
\setlength{\tabcolsep}{1pt}
\small
 \begin{tabular}{cccccccc}
  \hline
  longitude& latitude & tilt angle& height&ratio & half angle & STA time & SOHO time \\
  ($~^\circ~$) & ($~^\circ~$) & ($~^\circ~$) & ($R_s$) & & ($~^\circ~$) & (yyyy/mm/dd hh:mm) & (yyyy/mm/dd hh:mm) \\
  \hline
  90 & -10 & 45 & 2.5 & 0.6 & 60 & 2017/09/10 16:00~ & ~2017/09/10 16:00\\
  90 & -10 & 45 & 3.8 & 0.6 & 60 & 2017/09/10 16:05~ & ~2017/09/10 16:00\\
  90 & -10 & 45 & 5.6 & 0.6 & 60 & 2017/09/10 16:10~ & ~2017/09/10 16:12\\
  90 & -10 & 45 & 13.0 & 0.6 & 60 & 2017/09/10 16:24~ & ~2017/09/10 16:42\\
  90 & -10 & 45 & 15.5 & 0.6 & 60 & 2017/09/10 16:39~ & ~2017/09/10 16:54\\
  \hline
\end{tabular}
\ec
\end{table}

\begin{table}
\bc
\begin{minipage}[]{110mm}
\caption[]{Spheroid model parameters of shock at different time\label{tab:Shockfit}}\end{minipage}
\setlength{\tabcolsep}{1pt}
\small
 \begin{tabular}{ccccclcc}
  \hline
  longitude& latitude & tilt angle& e & kappa & height & STA time & SOHO time \\
  ($~^\circ~$) & ($~^\circ~$) & ($~^\circ~$) & & & $R_s$ & (yyyy/mm/dd hh:mm) & (yyyy/mm/dd hh:mm) \\
  \hline
  95 & -8  & 0 & -0.5 & 0.85 & 2.3+0.5 & 2017/09/10 16:00~ & ~2017/09/10 16:00\\
  95 & -8  & 0 & -0.5 & 0.85 & 3.5+0.5 & 2017/09/10 16:05~ & ~2017/09/10 16:00\\
  95 & -10 & 0 & -0.5 & 0.85 & 5.8+0.15 & 2017/09/10 16:10~ & ~2017/09/10 16:12\\
  95 & -10 & 0 & -0.5 & 0.85 & 13.7 & 2017/09/10 16:39~ & ~2017/09/10 16:42\\
  95 & -10 & 0 & -0.5 & 0.85 & 16.5 & 2017/09/10 16:54~ & ~2017/09/10 16:54\\
  \hline
\end{tabular}
\ec
\end{table}

\subsection{Release times}
\label{sect:release}

The relationship between the the kinetic energy and the rigidity, and the relationship between the speed and kinetic energy of a particle can be described by the two formulas (Le et al. 2006) listed below
\begin{equation}\label{eq:03}
E_k=\frac{-2E_0+\sqrt{(E_0)^2+(ZeR)^2}}{2}, ~~~~ v=\frac{\sqrt{E_k^2c^2+2E_km_0c^4}}{E_k+E_0}
\end{equation}
where $R$ is particle rigidity, $Z$ is particle charge number and $e$ is the charge of a electron, $E_k$ is the particle kinetic energy, while $E_0$ is the particle rest energy. $c$ is the light speed and $m_0$ is the rest mass of particle.

According to formula (\ref{eq:03}), we can obtain the speed of particle with kinetic energy $E_k$. When the path length traveled by particles is known, the solar release times of protons with different energies can be derived according to formula (\ref{eq:02}). The time information for particles with different energies is listed in Table \ref{tab:SEPtime}.

\begin{table}
\bc
\begin{minipage}[]{110mm}
\caption[]{The characteristic Times of the SEP event on 2017 September 10\label{tab:SEPtime}}\end{minipage}
\setlength{\tabcolsep}{1pt}
\small
 \begin{tabular}{ccccc}
  \hline
  Category& Onset time & Peak time& Traveling time & SRT \\
    &1AU(UT)&1AU(UT)&(minutes)&Sun(ST)\\
  \hline
  SXR(1-8~\AA)&15:35&16:06&8.33&15:26:40$^a$\\
    & & & &15:57:40$^b$\\
  %\hline
  HXR(100-300 keV)& 15:56&16:12&8.33&15:47:40\\
  %\hline
  HXR(300-1000 keV)& 15:59:30&&8.33&15:51:10\\
  %\hline
  Type III-l burst&15:45& &8.33&15:36:40\\
  %\hline
  Type II burst&$>$16:03& &8.33&$>$15:54:40\\
  %\hline
  CME entered LASCO C2&16:00& &8.33&15:51:40\\

   & & &\multicolumn{2}{c}{Path length =1.7 AU}\\

  Electrons(38-53 keV) & 16:15$\pm$5 &  & 38.7 & 15:36:18$\pm$5 \\

  Electrons(173-315 keV) & 16:10$\pm$5 &  & 21.3 & 15:48:42$\pm$5 \\

  Protons (E$>$30 MeV) & 16:20$\pm$5 & &52.2 & 15:28:48$\pm$5 \\

  Protons (E$>$50 MeV) & 16:20$\pm$5 &  &45.0 & 15:35$\pm$5 \\

  Protons (E$>$100 MeV) & 16:20$\pm$5 &  &33.0 & 15:47$\pm$5 \\

  Protons (E$>$700 MeV) & 16:15$\pm$5 &  &17.3 & 15:54:42$\pm$5 \\

  FSMT NM (1 GV) & 16:12$\pm$1 &  &19.4  &15:52:36$\pm$1 \\

  MGDN NM (2.09 GV) & 16:20$\pm$1 &  &15.5 & 16:04:30$\pm$1 \\

  APTY NM (1 GV) & 16:34$\pm$1 &  &19.4  &16:14:36$\pm$1 \\

  OULU NM (1 GV) & 16:40$\pm$1 &  &19.4  &16:20:36$\pm$1 \\

  SOPO NM (1 GV) & 16:42$\pm$1 & &19.4 & 16:22:36$\pm$1 \\
  \hline
\end{tabular}
\ec
$^a$ indicates the solar time for start time of the flare, $^b$ indicates the solar time for peak time of the flare
\end{table}

\section{Discussion and Summary}
\label{sect:discussion}

The velocity dispersion analysis (VDA) has three assumptions. The first one is that particles with different energies will be released at the same time in the Sun or near the Sun. The second one is that the path lengths for protons with different energies traveling from the Sun to the Earth are the same. The third one is that the pitch angles of first arriving particles are zero, namely that the propagation of particles with different energies is scatter free from the Sun to the Earth. According to VDA, the release times of RSPs seemly support that the RSPs were accelerated by the CME-driven shocks (Reames 2009a). However, statistical results show that for the SEP events with source locations in the well connected region, $E\ge30$ MeV protons may be mainly accelerated by the concurrent flares (Le et al. 2017a, 2017b), suggesting that the real path lengths traveled by particles from the Sun to the Earth should be longer than those obtained by VDA. According to the analysis of the radio emissions, Cane (2003) believed that interplanetary scattering must be occurring, implying that path length traveled by particles from the Sun to the Earth should be longer than the one obtained by VDA.

Wang and Qin (2015) suggested that VDA is only valid with impulsive source duration, low background and weak scattering in the interplanetary space or fast diffusion in solar atmosphere. The low background means that background level is below 0.01\% of the peak intensities of the flux (Wang and Qin, 2015). The background levels for both $E>10$ MeV is about 1\% of the peak intensities of the flux in the large SEP event occurred on 2017 September 10, and the situation for $E>100$ MeV protons is the same as for $E>10$ MeV protons, suggesting that background is middle and the real path length traveled by particles from the Sun to the Earth should be much longer than the one obtained by VDA (Wang and Qin. 2015).

It is more commonly accepted that perpendicular shock has stronger ability in accelerating particles than parallel shock(e.g., Tylka et al. 2005; Reames et al. 2012) although some researchers argued that parallel shock stronger ability in accelerating protons than perpendicular shock (e.g., Lee 2005; Zank et al. 2006). Very recently, a simulation study made by Qin et al. (2018) suggested that strong acceleration of particles at perpendicular shock is more significant than that at parallel shock. The source location of the SEP event of 2017 September 10 is S06W88, suggesting that only the particles accelerated by the shock driven by east flank of the CME can be observed by GOES spacecraft. The shock driven by the CME east flank is usually a parallel shock, which is often a weak shock. This kind of shock can hardly accelerate the protons to relativistic energy, implying that RSPs are more likely accelerated by the concurrent flare.

Type II radio emission occurs at the local plasma frequency and/or its harmonic, so the frequency of emission is indicative of the heliocentric distance at which the radio emission originates. Type II bursts are narrow band radio emissions typically drifting downward in frequency. Drift rates of type II bursts is the indicative of the speed of the shock driven by the associated CME. We can see from Figure 2 that the drift rates of type II bursts during the early phase is very slow, indicating that the speed of shock driven by east flank of the CME is very slow at the early phase, suggesting that the shock driven by CME east flank is really a weak shock. The enhancement in the flux of $E>10$ MeV protons was very small when the shock passed through the Earth shown in Figure 3, indicating that the shock was a very weak shock when it reached the Earth. The shock should be driven by the far east flank of the CME when the shock passed through the Earth, this may be the reason why the shock is so weak. To be noticed that the fitted speed of CME by GCS model is the speed of CME nose, not the speed of CME east flank.

\begin{figure}
   \centering
  \includegraphics[width=10cm, angle=90]{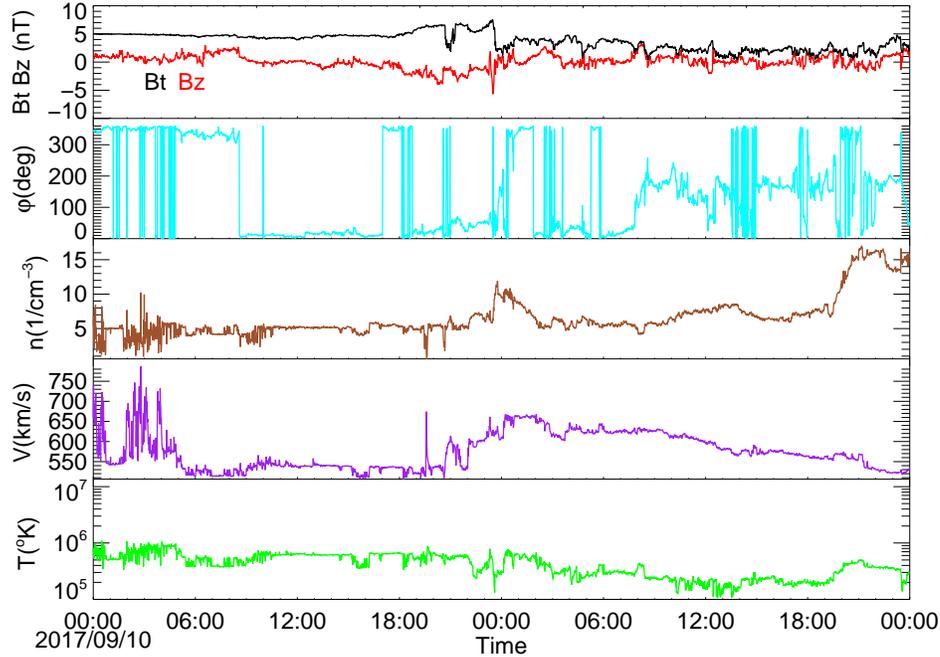}
   \caption{Solar wind parameters during 10-11, September 2017 observed by DSCOVR. From top to bottom, it indicates interplanetary magnetic field (IMF) with black line for total IMF and red line for z-component of IMF, azimuth $\phi$, proton densities, solar wind speed and proton temperature respectively.}
   \label{fig:SolarWind}
\end{figure}

Near relativistic electrons can be efficiently accelerated by perpendicular shock (e.g., Carley et al. 2013; Guo and Giacalone 2010; Kong et al. 2016). However, the shock driven by CME east flank is a parallel shock, which is not an efficient shock to accelerate electrons, implying that the near relativistic electrons in the SEP event of 2017 September 10 may be accelerated by the concurrent flare. Solar wind speed is about 550 km/s before the SEP event, which is shown in Figure \ref{fig:SolarWind}. The length of nominal park spiral line linked the Sun and the Earth is about 1.3 AU and the estimated longitude of the foot point on the Sun of the park spiral line is about W45 for solar wind speed around 550 km/s. Because the source location of the SEP event is S08W88, it is evident that SEPs observed by spacecraft GOES not only goes along the park spiral line, but also must have lateral propagation.

The source locations of two GLE events that occurred on 1982 November 26 and 1982 December 7 are S12W87 and S19W86 respectively. The path lengths for particles traveling from the Sun to the Earth inferred by VDA are 1.96 AU and 1.78 AU respectively for the two GLE events that occurred on 1982 November 26 and 1982 December 7 (Reames 2009b). Based on the assumption of scatter free, the path length traveled by particles from the Sun to the Earth inferred by VDA should be the shortest one, implying that real path lengths should be longer than 1.96 AU and 1.78 AU respectively for the two GLE events. The longitude of source location of SEP event that occurred on 2017 September 10 is W88, which is closer to the west limb of the Sun than those of the two SEP events that occurred on 1982 November 26 and 1982 December 7, suggesting that the path length traveled by protons from the Sun to the Earth for the SEP event of 2017 September 10 should be longer than 1.7 AU.

According to the path length traveled by particles and the onset times for particles with different energy, the derived solar release times for particles with different energies are different, which can be seen from Table \ref{tab:SEPtime}. We can also see from Table \ref{tab:SEPtime} that solar release times of $E>30$ MeV, $50$ MeV and $100$ MeV occurred during the impulsive phase of the X8.2 flare. In addition, the solar release time of $E>100$ MeV protons is earlier than the solar time of Type II radio burst, and the solar release times of $E>30$ MeV and $E>50$ MeV protons are much earlier than the solar time of Type II radio burst, suggesting that the first arriving $E>30$ MeV, $E>50$ MeV and $E>100$ MeV protons should be accelerated by the concurrent flare.

The onset times for both $E>$30 MeV protons and $E>$100 MeV protons are 16:20 UT $\pm$5 minutes, indicating that the time difference between the onset times for $E>$30 MeV protons and $E>$100 MeV protons should not exceed 10 minutes. However, the derived solar release time of $E>$30 MeV protons is 19.2 minutes earlier than that of $E>$100 MeV protons, suggesting that the release time for E$>$30 MeV protons is at least 9.2 minutes earlier than that for E$>$100 MeV protons, namely that protons with lower energy may leave the Sun earlier than those with higher energy.

The earliest onset time of RSPs is observed by FMST NM. The derived solar release time of RSPs observed by FMST NM is 15:52$\pm$1 min ST, which is earlier than the start time of type II radio burst. The time that CME just entered the LASCO C2 view field is consistent with the time when first arriving RSPs reached the Earth and any interaction of the CME with the structures higher in the corona could not have happened and could not played a role in the acceleration of the first arriving high energy proton. We can also see from Table \ref{tab:SEPtime} that the onset times of RSPs observed by several NMs were later than the peak time of SXR flux and also latter than the first peak time of 100-300 keV HXR flux, suggesting that the protons accelerated by the concurrent flare may be further accelerated by the CME-driven shock to higher energy.

Solar release times estimated for different energy protons are based on the assumption that the path length traveled by protons is the same as that traveled by electrons. However, the path length traveled by protons is usually longer than that traveled by electrons. As a result, the solar release times for protons with different energies should be earlier than those listed in Table 3. This will further support that the first arriving protons are accelerated by concurrent flare.

According to the data analyses and the discussion above, the results can be summarize as below:

(1) The cosmic rays are highly anisotropy at the early phase and few protons have energy greater than 1.64 GeV. Both the first arriving RSPs and non-RSPs may be accelerated by the concurrent flare, and first near relativistic electrons may also be accelerated by concurrent flares.

(2) The release times of protons with different energies may be different. The protons with lower energy may release earlier than those with higher energy.

(3) The estimated averaged CME speed was about 2808.8 km/s in LASOC C2 and C3 view field, while
CME-driven shock speed was about 2933.5 km/s in LASOC C2 and C3 view field. The protons accelerated by the concurrent flare may be further accelerated by the CME-driven shock.

(4) The interaction of CME with the structures higher in the corona could not have played a role in the acceleration of the first arriving high energy protons.

\normalem
\begin{acknowledgements}

We are very grateful to the anonymous referee for her/his reviewing of the paper and for helpful suggestions. We also thank Scientific Editor for his useful comments. We are grateful to SOHO, STEREO, GOES, ACE, FERMI and NMDB for making their data available on line. This work was jointly funded by the National Natural Science Foundation of China (grants 41674166, 41074132, 41274193, 41304144), the National Standard Research Program (Grant 200710123).

\end{acknowledgements}

\end{document}